%% file: ms.tex
\documentclass{article}
\pdfoutput=1 
\usepackage{arxiv}
\usepackage[utf8]{inputenc} 
\usepackage[T1]{fontenc}    
\usepackage{hyperref}       
\usepackage{url}            
\usepackage{booktabs}       
\usepackage{amsfonts}       
\usepackage{nicefrac}       
\usepackage{microtype}      
\usepackage{amsmath,amssymb,amsfonts}
\usepackage{longtable}
\usepackage{graphicx}
\usepackage{subfiles}
\usepackage{subcaption}

\newcommand{\ket}[1]{|#1\rangle}

\title{Quantum gene regulatory networks}
\date{}
\author{
    Cristhian~Roman-Vicharra\\
    Department of Veterinary Integrative Biosciences\\
    Texas A\&M University\\
    College Station, TX 77843\\
    \texttt{cristhianroman@tamu.edu} \\
\And
    James J.~Cai
    \thanks{Correspondence}\\
    Department of Veterinary Integrative Biosciences\\
    Department of Electrical and Computer Engineering\\
    Texas A\&M University\\
    College Station, TX 77843\\
    \texttt{jcai@tamu.edu} \\
}

\renewcommand{\headeright}{}
\renewcommand{\undertitle}{}
\renewcommand{\shorttitle}{Quantum gene regulatory networks}

\hypersetup{
pdftitle={Quantum gene regulatory networks},
pdfauthor={Cristhian~Roman-Vicharra, James J.~Cai},
pdfsubject={In this work, we present a quantum circuit model for inferring gene regulatory
networks (GRNs). The model is based on the idea of using qubit-qubit
entanglement to simulate interactions between genes. We provide preliminary
results that suggest our quantum GRN modeling method is competitive and warrants
further investigation.}
}

\begin{document}
\maketitle

\begin{abstract}
In this work, we present a quantum circuit model for inferring gene regulatory
networks (GRNs). The model is based on the idea of using qubit-qubit
entanglement to simulate interactions between genes. We provide preliminary
results that suggest our quantum GRN modeling method is competitive and warrants
further investigation. Specifically, we present the results derived from the
single-cell transcriptomic data of human cell lines, focusing on genes in
involving innate immunity regulation. We demonstrate that our quantum circuit
model can be used to predict the presence or absence of regulatory interactions
between genes and estimate the strength and direction of the interactions,
setting the stage for further investigations on how quantum computing finds
applications in data-driven life sciences and, more importantly, to invite
exploration of quantum algorithm design that takes advantage of the single-cell
data. The application of quantum computing on single-cell transcriptomic data
likewise contributes to a novel understanding of GRNs, given that the
relationship between fully interconnected genes can be approached more
effectively by quantum modeling than by statistical correlations.
\end{abstract}

\paragraph{Summary}
\subfile{./summary.tex}

\section{Introduction}
\subfile{./introduction.tex}

\section{Methods}
\subfile{./methods/quantum_computing_theory.tex}

\subfile{./methods/qscgrn_model.tex}
\subfile{./methods/quantum_classical_frame.tex}
\subfile{./methods/data_sets.tex}

\section{Results}
\subfile{./results/real_data.tex}

\section{Discussion}
\subfile{./discussion.tex}

\section{Acknowledgments}
This work was supported by the DoD grant GW200026 for JJC.

\section{Data availability}
The scRNA-seq data analyzed during the current study is available in the NCBI
GEO database with the accession numbers GSE126321 and GSE158275. The processed
data and the source code implementation of the qscGRN package are provided
in the GitHub repository at \url{https://github.com/cailab-tamu/QuantumGRN/}.
The repository also includes tutorials written in Python language.

\section{Author Contributor}
Conceptualization, JJC; methodology, CR and JJC; implementation of the
software, CR; formal analysis, CR and JJC; writing and editing, CR and JJC;
supervision, JJC. All authors reviewed and contributed to the manuscript.

\section{Competing Interest}
The authors declare no competing financial or non-financial interest.
\newpage

\bibliographystyle{unsrt}
\bibliography{./references.bib}
\end{document}

%% file: summary.tex
Quantum computing holds the promise to achieve certain types of computation
that would otherwise be unachievable by classical computers. The advent in
the development of quantum algorithms has enabled a variety of applications
in chemistry, finance, and cryptography. Here we introduce a parameterized
quantum circuit modeling method for constructing gene regulatory networks
(GRNs) using data from single-cell RNA sequencing (scRNA-seq). In the circuit,
each qubit represents a gene, and qubits are entangled to simulate the
interaction between genes. The strength of interactions is estimated using
the rotation angle of controlled unitary gates between qubits after fitting the
scRNA-seq expression matrix data. We applied our quantum single-cell GRN (qscGRN)
model to real scRNA-seq data obtained from human lymphoblastoid cell lines and
demonstrated its usage in recovering known and detecting novel regulatory
relationships between genes in the nuclear factor-kappa B (NF-$\kappa$B)
signaling pathway. Our quantum circuit model enables the modeling of vast
feature space occupied by cells in different transcriptionally activating
states, simultaneously tracking activities of thousands of interacting genes
and constructing more realistic single-cell GRNs without relying on
statistical correlation or regression. We anticipate that quantum computing
algorithms based on our circuit model will find more applications in
data-driven life sciences, paving the way for the future of predictive
biology and precision medicine.

%% file: introduction.tex
A gene regulatory network (GRN) defines the ensemble of regulatory relationships
between genes in a biological system. Inferring GRNs is a powerful approach for
studying molecular mechanism of transcriptional regulation and the function of
genes in processes of cellular activities \cite{Huynh-Thu2019, HECKER200986}. A
GRN is often represented as a graph—which can be signed, directed, and
weighted—representing the relationships between transcription factors or
regulators and their target geneswhose expression level is controlled. However,
because gene regulation inside cells is difficult to observe, indirect
measurements of intracellular expression are often used as a proxy, and the
statistical dependencies are used to infer real regulatory relationships between
genes. Thus, the power of different methods for GRN inference depends on the
types of computational algorithms and the resolution of the expression data
\cite{chen2018evaluating, pratapa2020benchmarking, diaz2022gaining}.

Single-cell technologies, which have recently been developed and improved, open
up opportunities for studying biology at unprecedented resolution and scale.
Single-cell RNA sequencing (scRNA-seq) allows us to measure gene expression in
individual cells for thousands of cells in a single experiment
\cite{zheng2017massively}. The GRN modeling can adopt scRNA-seq technology and
leverage the unprecedented information from the sheer number of cells to improve
inference power \cite{chan2017gene}. The use of such data would allow us to
learn better and more detailed network models, which will also help us better
understand the mechanics behind cellular operations. A plethora of computational
methods for inferring GRNs have been developed for either population-level or
single cell-level gene expression data. These methods apply statistical
approaches to identify likely regulatory relationships between genes based on
their expression patterns. These different methods are based on correlation and
partial correlation \cite{kim2015ppcor}, information theory \cite{chan2017gene},
regression \cite{osorio2020sctenifoldnet, huynh2010inferring}, Gaussian
graphical model \cite{kotiang2020probabilistic}, Bayesian and Boolean networks
\cite{lahdesmaki2006relationships, friedman2000using,
shmulevich2002probabilistic}, and many others. Each method has its own set of
assumptions and limitations,which are not always stated explicitly.

In recent years, quantum computing has become an emerging technology and an
intense field of research constantly seeking applications
\cite{GYONGYOSI201951}. Researchers have developed quantum algorithms with
applications in areas such as finance \cite{qcomputingfinance}, cryptography
\cite{qcomputingcryptography}, machine learning \cite{qcomputingml}, drug
discovery \cite{koes2018pharmit}, chemistry, and material science
\cite{bauer2020quantum}. Theoretically, a speedup is expected in certain types
of computation using quantum algorithms versus classical algorithms because a
quantum computer takes advantage of superposition and entanglement phenomena
during the computation \cite{RevModPhys.94.015004, huang2022quantum}. The most
iconic quantum algorithm is Shor’s \cite{shorfact} for the factorization of
large numbers, which can break the Rivest Shamir Adleman encryption
\cite{rsaencp}. Due to the potential of quantum computing, the current approach
to scRNA-seq analysis and GRN inference may be rethought.

In this work, we introduce a quantum single-cell GRN (qscGRN) modeling method,
which is based on a parameterized quantum circuit and uses the quantum framework
to recover biological GRNs from scRNA-seq data. In the qscGRN model, a gene is
represented using a qubit, and the structure is divided into 2 types of layers:
the encoder layer that translates the scRNA-seq data into a superposition state
and the regulation layers that entangle qubits and model gene-gene interactions
in the quantum framework. In contrast to the correlation-based inference 
methods, the qscGRN model maps the binarized expression values onto a large 
vector space, known as Hilbert space, making full use of the cell information
in the scRNA-seq data. Thus, a large number of cells in the scRNA-seq data is
important because it improves the mapping of biological information in a
superposition state. In addition, parameterization in the qscGRN model allows
the gene-gene relationships to be inferred all at once by fitting the
superposition state probabilities onto the distribution observed in the
scRNA-seq data.

A quantum-classical framework for optimizing the qscGRN model is also
introduced. The classical component of our framework uses the Laplace smoothing
\cite{peng2004augmenting} and the gradient descent algorithm
\cite{gradientdescentov} to perform optimization by minimizing a
Kullback-Leibler (KL) divergence \cite{kldivergence} as a loss function.
Finally, we used the quantum-classical framework on a real scRNA-seq data set
\cite{osorio2019single, sorelle2021single} to show that gene-gene interactions
can be modeled using quantum computing, and the structure of such as previously
published GRN \cite{roy2019regulatory, sciammas2011incoherent} can be recovered
from the parameter-optimized quantum circuit.

%% file: methods/quantum_computing_theory.tex
\subsection{Quantum computing theory}
We first introduce the basic, broad-audience background of quantum computing
necessary for this work. Classical computers manage information processing using
bits for storage, computation, and communication \cite{archquantumcomp}. A bit
is the unit of information being $0$ or $1$, also represented in Dirac notation
\cite{dirac_1939} as $\ket{0} {=} \begin{pmatrix} 1 & 0 \end{pmatrix}^\intercal$ or
$\ket{1} {=} \begin{pmatrix} 0 & 1 \end{pmatrix}^\intercal$ respectively
\cite{beginquantumc}. In quantum computing, a qubit is the unit of information
represented as $\ket{\psi} {=} \begin{pmatrix} c_0 & c_1 \end{pmatrix}^\intercal {=}
c_0\ket{0} + c_1\ket{1}$, where $\ket{\psi}$ is the quantum state in the
superposition of $\ket{0}$ and $\ket{1}$ basis in a $1$-dimensional Hilbert
space, and $c_0$, $c_1$ are complex numbers. The state of a quantum system is
described by a unit vector in the Hilbert space; therefore, the square modulus
sum $|c_0|^2+|c_1|^2$ is equal to $1$. In quantum mechanics, the measure of
$\ket{\psi}$ results in $0$ with a probability to be observed of $|c_0|^2$, and
$1$ of $|c_1|^2$. Thus, the probability of measuring a basis is the squared
modulus of the associated complex number.

\begin{figure}[htbp]
\centerline{
    \includegraphics[width=0.8\textwidth,trim=0 10 0 3, clip]{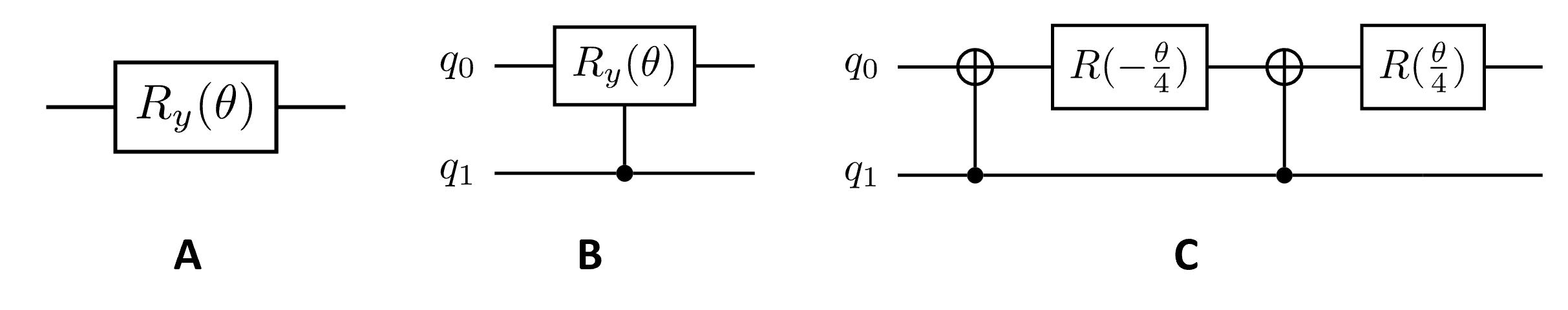}
}
\caption{Schematic representations of $R_y$ gate, controlled-$R_y$ gate, and its
decomposition. (\textbf{A}) A schematic representation of the $R_y$ gate with a
rotation parameter $\theta$. In the representation, the input register is on the
left, and the output is on the right. (\textbf{B}) A schematic representation of
the controlled-$R_y$ gate with a rotation parameter $\theta$. In the
representation, the control register is the qubit labeled $q_1$, and the target
is labeled $q_0$. The operation is performed on the target register when the
control register $q_1$ is $1$, and the rotation is not performed otherwise.
(\textbf{C}) The schematic representation for the decomposition of a
controlled-$R_y$ gate with a rotation parameter $\theta$. The sequence consists
of 2 rotation gates $R(\frac{\theta}{4})$, $R(-\frac{\theta}{4})$ gates and 2
c-$NOT$ gates. The operation in the target register is equal to an $R_y(\theta)$
gate when the control register $q_1$ is 1, and no operation is performed
otherwise.}
\label{figure1}
\end{figure}

Single-qubit gates that are widely used include the $NOT$ gate, Hadamard
gate, Pauli gates $X$, $Y$ and $Z$, phase shift gates, and
parameterized rotation gates $R_x$, $R_y$ and $R_z$. The Hadamard
gate—represented as $H$ gate—is frequently used in various quantum algorithms
and is defined as $\tfrac{1}{\sqrt{2}} \big(\begin{smallmatrix}
  1 & 1\\
  1 & -1
\end{smallmatrix}\big)$. The $H$ gate maps the basis state $\ket{0}$ to
$H\ket{0} {=} \tfrac{1}{\sqrt{2}} \big(\begin{smallmatrix}
  1 & 1\\
  1 & -1
\end{smallmatrix}\big) \big(\begin{smallmatrix}
  1\\
  0
\end{smallmatrix}\big) {=} \tfrac{1}{\sqrt{2}} \big(\begin{smallmatrix}
  1\\
  1
\end{smallmatrix}\big) {=} \tfrac{\ket{0} + \ket{1}}{\sqrt{2}}$ and $\ket{1}$ to
$\tfrac{\ket{0} - \ket{1}}{\sqrt{2}}$, creating a superposition of the basis
states. The measurement of the quantum state $\tfrac{\ket{0} +
\ket{1}}{\sqrt{2}}$ results in observing the basis state $\ket{0}$ with a
probability of $0.5$ and $\ket{1}$ with $0.5$. Furthermore, the rotation gate
$R_y$ is a single-qubit operation (FIGURE \ref{figure1}A) based on the
exponentiation of the Pauli gate \emph{Y} using a rotation parameter $\theta$
and is defined as $$
R_y(\theta) \equiv e^{\frac{-i\theta Y}{2}} = \cos{\frac{\theta}{2}} I -
    i \sin{\frac{\theta}{2}} Y = \begin{pmatrix}
        \cos{\frac{\theta}{2}} & -\sin{\frac{\theta}{2}} \\
        \sin{\frac{\theta}{2}} & \cos{\frac{\theta}{2}}
    \end{pmatrix},
$$ where \emph{Y} is a Pauli operation defined as $\big(\begin{smallmatrix}
  0 & -i\\
  i & 0
\end{smallmatrix}\big)$ and $I$ is the identity matrix. The $R_y$ gate maps the
basis state $\ket{0}$ to a superposition state $R_y(\theta)\ket{0} {=}
\cos{\frac{\theta}{2}} \ket{0} + \sin{\frac{\theta}{2}} \ket{1}$.

In addition, the controlled gate is a 2-qubit gate—which applies an operation
on a target qubit when the control qubit is in state $\ket{1}$. The operation is
typically a single-qubit gate such as $R_y$ gate. Thus, a controlled-$R_y$ gate,
represented as c-$R_y$ gate, is defined as $$
\text{c-}R_{y,1,0}(\theta) = \begin{pmatrix}
    1 & 0 & 0 & 0\\
    0 & 1 & 0 & 0\\
    0 & 0 & \cos{\frac{\theta}{2}} & -\sin{\frac{\theta}{2}}\\
    0 & 0 & \sin{\frac{\theta}{2}} & \cos{\frac{\theta}{2}}
\end{pmatrix},
$$ where the control register is the qubit labeled $q_1$ and the target is $q_0$
(FIGURE \ref{figure1}B). In the case where the control register is the qubit
labeled $q_0$ and the target is $q_1$, the c-$R_y$ gate is defined as $$
\text{c-}R_{y,1,0}(\theta) = \begin{pmatrix}
    1 & 0 & 0 & 0\\
    0 & \cos{\frac{\theta}{2}} & 0 & -\sin{\frac{\theta}{2}}\\
    0 & 0 & 1 & 0\\
    0 & \sin{\frac{\theta}{2}} & 0 & \cos{\frac{\theta}{2}}
\end{pmatrix}.
$$

TABLE \ref{table1} shows the mapping of the basis states in a 2-dimension Hilbert space
when using a c-$R_{y,1,0}(\theta)$ gate. The basis states $\ket{00}$ and
$\ket{01}$ have the first bit as $0$, thus the operation $R_y$ is not performed.
On the other hand, the basis states $\ket{10}$ and $\ket{11}$ have the first bit
as $1$, thus the operation $R_y$ is performed in the second bit.

\begin{longtable}[c]{ c | c }
\caption{The controlled-$R_{y,1,0}(\theta)$ gate mapping of the basis state in
a 2-dimension Hilbert space. The $R_y$ operation in the second bit is performed
when the first bit is $1$, and the second bit does not change otherwise.
\label{controlledry}}\\
\hline
Basis state $\ket{\mathbf{x}}$ & c-$R_{y,1,0}(\theta)$\\
\hline
\endfirsthead
\hline
\endfoot

$\ket{00}$ & $\ket{00}$\\
$\ket{01}$ & $\ket{01}$\\
$\ket{10}$ & $\cos{\frac{\theta}{2}} \ket{10} + \sin{\frac{\theta}{2}} \ket{11}$\\
$\ket{11}$ & $-\sin{\frac{\theta}{2}} \ket{10} + \cos{\frac{\theta}{2}} \ket{11}$
\label{table1}
\end{longtable}

Generally, a controlled-$U$ gate can be decomposed into single-qubit gates such
as phase shift, c-$NOT$ and rotation gates, where $U$ is a single qubit gate.
Thus, the c-$R_y$ gate is decomposable only into 2 single rotation gates,
represented as $R(c) {=} \big(\begin{smallmatrix}
    \cos{c} & -\sin{c} \\
    \sin{c} & \cos{c}
\end{smallmatrix}\big)$, and 2 c-$NOT$ gates because there is no phase shift
operation. FIGURE \ref{figure1}C shows the decomposition of a c-$R_{y,1,0}$ gate
into $R(\frac{\theta}{4})$, $R(-\frac{\theta}{4})$ gates and 2 c-$NOT$ gates,
where the control register is $q_1$, and target is $q_0$. In other words,
the effects of the rotation gates sum up $-\frac{\theta}{2}$ when the control
qubit is $1$ and cancel out each other otherwise.

%% file: methods/qscgrn_model.tex
\subsection{The qscGRN model: a parameterized quantum circuit}
In classical computation, a circuit is a model composed of a sequence of
instructions (NOT, AND, OR classical gates) that are not necessarily reversible.
In the classical circuit, the input bits flow through the sequence of
instructions computing output bits for a certain task \cite{lee2015past}.
Similarly, a quantum circuit is a model consisting of a sequence of quantum
gates that perform operations on the qubits \cite{generalcircuit}. A quantum
circuit that is running an algorithm is usually initialized to $\ket{0}_n$,
which means a string of $n$ bits of all zeros, and then put into a superposition
state using $H^{\otimes n}$ transformation—which means an H gate on each
qubit—allowing all possible inputs to be tested \cite{quantumalgorithms}. Then,
the register flows through a sequence of quantum gates, and the output register
is measured and decoded to interpret the result of the algorithm.

\begin{figure}[htbp]
\centerline{
    \includegraphics[width=0.8\textwidth,trim=0 10 0 3, clip]{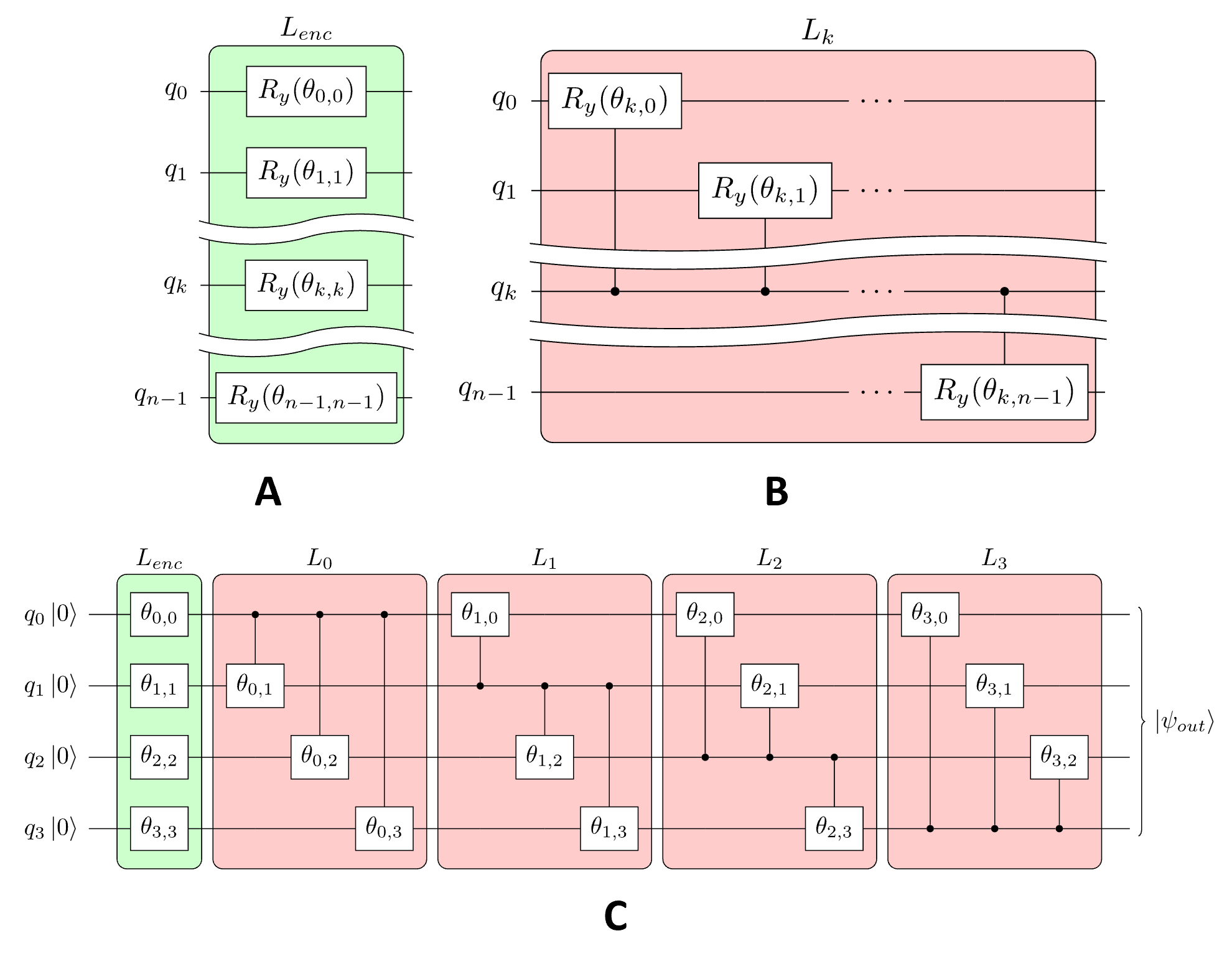}
}
\caption{The quantum single-cell gene regulatory network (qscGRN) model.
(\textbf{A}) The encoder layer $L_{enc}$ for a qscGRN model that has $n$
qubits, represented with a green background. The $k$th qubit has an $R_y$
gate with a parameter $\theta_{k,k}$—which is a diagonal element of the
adjacency matrix in the qscGRN. (\textbf{B}) The regulation layer $L_k$ for
a qscGRN model that has $n$ qubits, represented with a red background. The
$k$th qubit uses a c-$R_y$ gate with a parameter $\theta_{k,p}$—which is a
non-diagonal element of the adjacency matrix in the qscGRN—to connect to
the $p$th qubit, where $0 \leq p < n$ and $p \neq k$, thus the $k$th qubit
is fully connected to other qubits. (\textbf{C}) A schematic representation
of a qscGRN model that consists of 4 qubits. The $L_{enc}$ layer consists of
4 $R_y$ gates that put the input register into a superposition state.
The $L_k$ layer consists of c-$R_y$ gates that connect the $k$th qubit to
the others. Thus, a 4-qubit qscGRN model uses an $L_{enc}$ with 4 parameters
and $L_0$, $L_1$, $L_2$ and $L_3$ with 3 parameters for each layer. The
collection of parameters $\boldsymbol{\theta}$ is the matrix representation
of the 4-qubit qscGRN model.}
\label{figure2}
\end{figure}

Here, we introduce the quantum single-cell gene regulatory network (qscGRN)
model, that is a quantum circuit consisting of $n$ qubits, and models a
biological scGRN for $n$ genes in the framework of quantum computing. A qubit
in the qscGRN model represents a gene in the biological scGRN. The sequence of
gates is grouped into 2 types of layers: The encoder layer $L_{enc}$ consists
of $n$ $R_y$ gates that translate biological information (i.e., the state of
gene activity or the frequency of genes actively expressed among cells) onto a
superposition state. In the $L_{enc}$ layer, each qubit has a $R_y$ gate
(FIGURE \ref{figure2}A). The regulation layer $L_k$ consists of a sequence of
$n{-}1$ c-$R_y$ gates that have the $k$th qubit as control and a corresponding
target such that the $k$th qubit is fully connected to other qubits
(FIGURE \ref{figure2}B). In the $L_k$ layer, a c-$R_y$ gate—that has the $k$th
qubit as control and the $p$th qubit as the target—models the regulation
relationship in the corresponding gene-gene pair. In particular, the parameter
of a c-$R_y$ gate quantifies the strength and determines the type of
relationship between the $k$th and $p$th genes.

In FIGURE \ref{figure2}, we use the notation $\theta_{k,k}$ for the
parameter of the $R_y$ gate on the $k$th qubit in the $L_{enc}$ layer and
$\theta_{k,p}$ for the parameter on the c-$R_y$ gate in the $L_k$ layer that
has the $k$th and $p$th qubits as control and target, respectively. Formally,
the matrix representations for both layers are $$
L_{enc} = R_y(\theta_{n-1,n-1}) \otimes \cdots
    \otimes R_y(\theta_{1,1}) \otimes R_y(\theta_{0,0}),
$$ where the $\otimes$ operator is the tensor product, and $$
L_k = \prod_{i=0, i \neq k}^{n-1} R_{y,n}(\theta_{k,i})
    = R_{y,n}(\theta_{k,n-1}) \cdots R_{y,n}(\theta_{k,1})
    R_{y,n}(\theta_{k,0}),
$$ where $R_{y,n}(\theta_{k,i})$ denotes a c-$R_y$ gate with the $k$th qubit
as the control and $i$th qubit as the target in a $n$-qubit quantum circuit.
Also, the computation of the matrix representation is not commutative, which
means the order of the terms cannot be changed due to the operations needed
are matrix multiplication and tensor product.

The qscGRN model is initialized to $\ket{0}_n$ state, and then put into a
superposition state using a $L_{enc}$ layer. The gene-gene interactions are
then modeled using regulation layers $L_0$, $L_1$, $\cdots$, $L_{n-1}$. Thus,
the qscGRN model is a quantum circuit where each qubit is fully connected to
every other qubit and has a total of $n^2$ quantum gate parameters. Next, we
construct the matrix representation $\boldsymbol {\theta}$ of the qscGRN model
using the collection of parameters $\theta_{k,p}$ on the quantum gates, where
$0 \leq k$, $p < n$. Therefore, the matrix representation of the qscGRN model
is denoted as $$
\boldsymbol{\theta} = \begin{bmatrix}
    \theta_{0,0} & \theta_{0,1} & \cdots & \theta_{0,n-1} \\
    \theta_{1,0} & \theta_{1,1} & \cdots & \theta_{1,n-1} \\
    \vdots & \vdots & \ddots & \vdots \\
    \theta_{n-1,0} & \theta_{n-1,1} & \cdots & \theta_{n-1,n-1} \\
\end{bmatrix}
$$ where the diagonal elements belong to the $R_y$ gates in the $L_{enc}$ layer,
and the non-diagonal elements to the c-$R_y$ gates in the regulation layers
$L_0$, $L_1$, $\cdots$, $L_{n-1}$.

The output register $\ket{\psi}$ of the $n$-qubit qscGRN model, encodes the
gene-gene interactions in a superposition state according to the parameter
$\boldsymbol{\theta}$ and is formally defined as $$
\ket{\psi} = \left(\prod_{k=0}^{n-1} L_k\right) L_{enc} \ket{0}_n
    = L_{n-1} \cdots L_1 L_0 L_{enc} \ket{0}_n.
$$

FIGURE \ref{figure2}C shows the schematic representation of a qscGRN model
consisting of 4 qubits as an example for better interpretation of the equations.
The quantum gate name is not shown for simplicity but only the corresponding
parameter. According to the equations, the output register $\ket{\psi_{out}}$ in
FIGURE \ref{figure2}C is defined as $$
\ket{\psi_{out}} = L_3 L_2 L_1 L_0 L_{enc} \ket{0}_4,
$$ where $L_3$, $L_2$, $L_1$ and $L_0$ have 3 parameters each, and $L_{enc}$
has 4 parameters. Then, the matrix representation $\boldsymbol{\theta}$ of the
4-qubit qscGRN model is denoted as $$
\boldsymbol{\theta} = \begin{bmatrix}
    \theta_{0,0} & \theta_{0,1} & \theta_{0,2} & \theta_{0,3} \\
    \theta_{1,0} & \theta_{1,1} & \theta_{1,2} & \theta_{1,3} \\
    \theta_{2,0} & \theta_{2,1} & \theta_{2,2} & \theta_{2,3} \\
    \theta_{3,0} & \theta_{3,1} & \theta_{3,2} & \theta_{3,3}.
\end{bmatrix}
$$
Finally, we understand the matrix $\boldsymbol{\theta}$ as the adjacency matrix of the biological scGRN, which is a weighted directed fully connected network.

%% file: methods/quantum_classical_frame.tex
\subsection{Quantum-classical framework for optimization of the qscGRN model}
\subsubsection{Gene selection and binarization}
The input data of the workflow is a transformed scRNA-seq expression data
matrix $\boldsymbol{X}$ that has expression values for $m$ cells. The
transformation of the expression matrix can be done using, for example, Pearson
residuals \cite{lause2021analytic}. Then, we select $n$ genes from
$\boldsymbol{X}$ and binarize the expression values. The binarization is
achieved by applying an expression threshold of $0$ to the transformed
expression matrix, which means that expression values greater than $0$ are set
to $1$, and $0$ otherwise. The outcome of the binarization is saved to
$\boldsymbol{X^b}$, which is a matrix of dimension $n \times m$.

\subsubsection{Labeling and activation ratios (FIGURE \ref{figure3}A)}
\label{pobs}
A label is assigned for each cell in $\boldsymbol{X^b}$, such that the label is
a string composed of the binarized expression of the $n$ genes in a cell. In
other words, a label is the activation state of genes in a cell (colored in
light blue). Then, we compute the percentage of occurrences of each label in
the $m$ cells to obtain the observed distribution $p^{obs}$. The percentage of
label $\ket{0}_n$ in $p^{obs}$ is set to $0$, and the rest of the distribution
is rescaled to sum to $1$. The rationale for setting the $\ket{0}_n$
probability to $0$ is that cells with no expression levels are not informative
in the quantum framework due to dropout in the scRNA-seq experiment.
Furthermore, the activation ratio $act_k$ of the $k$th gene is defined as the
percentage of cells expressing that gene in $\boldsymbol{X^b}$ (colored in
light yellow). The $n$ genes in $\boldsymbol{X^b}$ are ordered decreasingly
by the activation ratio such that $\boldsymbol{X^b}$ has $n$ ordered rows
labeled as $g_0$, $g_1$, $\cdots$, $g_{n-1}$.

\subsubsection{Initialization of the parameter
\texorpdfstring{$\boldsymbol{\theta}$}{Theta} in the qscGRN model
(FIGURE \ref{figure3}B)}
\label{sectiontheta}
The parameters $\theta_{k,p}$ in the regulation layers $L_0$, $L_1$,
$\cdots$, $L_{n-1}$ are initialized to 0, where $0 \leq k, p < n$ and $k 
\neq p$. Besides, the parameters $\theta_{k,k}$ in the encoder layer $L_{enc}$
are initialized to $2 \arcsin{\sqrt{act_k}}$ corresponding to the $k$th gene,
where $0 \leq k < n$. Therefore, the initial parameter $\boldsymbol{\theta}$ is
represented as $$
\theta_{initial} = \begin{pmatrix}
    2 \arcsin{\sqrt{act_0}} & 0 & \cdots & 0 \\
    0 & 2 \arcsin{\sqrt{act_1}} & \cdots & 0 \\
    \vdots & \vdots & \ddots & \vdots \\
    0 & 0 & \cdots & 2 \arcsin{\sqrt{act_{n-1}}}
\end{pmatrix}.
$$

The rationale for the formula $2 \arcsin{\sqrt{act_k}}$ is that, independently
on each qubit, the probability of observing 1 is the activation ratio of the
corresponding gene after the $L_{enc}$ layer.

\subsubsection{Optimization of the parameter
\texorpdfstring{$\boldsymbol{\theta}$}{Theta} in the
qscGRN model (FIGURE \ref{figure3}C)}
We measure the output register $\ket{\psi_{out}}$ of the qscGRN model to
obtain the output distribution $p^{out}$ of observing the basis states. The
probability of the state $\ket{0}_n$ in $p^{out}$ is set to $0$, and the rest
of the distribution is rescaled to sum to $1$. Then, Laplace smoothing is used
to reshape $p^{obs}$ and $p^{out}$ to a different distribution $\hat{p}^{obs}$
and $\hat{p}^{out}$, respectively, thus handling the zero-probability problem
when computing the loss function. The smoothed distribution is computed as $$
\hat{p}^i = \frac{\#ocu^i + \alpha}{m + 2^n \alpha}
$$ where $i \in \{out,obs\}$ and $\alpha$ is the smoothing parameter which is
typically 1. $\#ocu^i$ is the number of occurrences in the distribution $p^i$.
In other words, $p^i {=} \tfrac{\#ocu^i}{m}$ is the original distribution.

The optimization of $\boldsymbol{\theta}$ is achieved by minimizing the loss
function to a threshold value of $2^n {\times} 10^{-6}$ using the gradient
descent algorithm with a learning rate $lr$ of 1. Otherwise, the optimization
is performed for pre-defined iterations $t$. The loss and error function are
defined as $$
L(\boldsymbol{\theta}) = \sum_{\mathbf{x} \in \{0,1\}^n} 
    \hat{p}^{out}_{\mathbf{x}} 
    \log{\left(\frac{\hat{p}^{out}_{\mathbf{x}}}
        {\hat{p}^{obs}_{\mathbf{x}}}\right)}
,$$ $$
E(\boldsymbol{\theta}) = \sum_{\mathbf{x} \in \{0,1\}^n}
    (p^{out}_{\mathbf{x}} -
        p^{obs}_{\mathbf{x}})^2,
$$ where $p^i_{\mathbf{x}}$ and $\hat{p}^i_{\mathbf{x}}$ are the
probability of the state $\mathbf{x}$ in the distributions,
$i \in \{out,obs\}$.

The parameters $\theta_{k,k}$ in the $L_{enc}$ layer are not trained during
optimization according to the assumption that these parameters encode a unique
binarized scRNA-seq matrix into the quantum framework. Thus, no training of the
parameters $\theta_{k,k}$ implies that the $L_{enc}$ layer encodes the same
biological information onto a superposition state in each iteration, making
the optimized parameter $\boldsymbol{\theta}$ meaningful from a biological
perspective.

\begin{figure}[htbp]
\centerline{
    \includegraphics[width=0.8\textwidth,trim=0 2 0 3, clip]{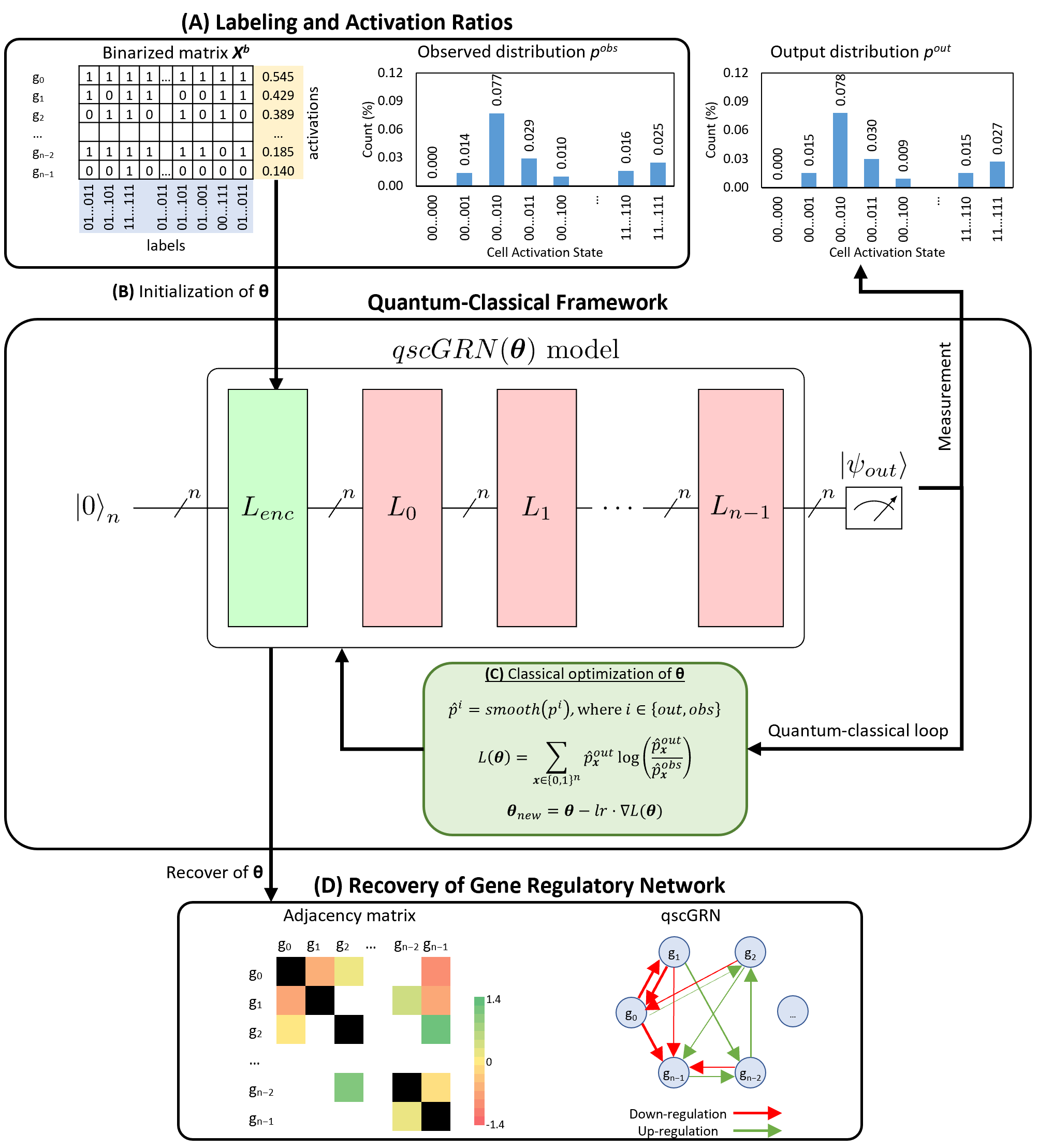}
}
\caption{The quantum-classical framework using the qscGRN model to infer the
corresponding biological scGRN. The input matrix $\boldsymbol{X^b}$ is the
binarized scRNA-seq data with n genes selected. (\textbf{A}) Labels are
assigned to each cell in $\boldsymbol{X^b}$ such that a label is the binarized
expression string for the $n$ genes in a cell. The observed distribution
$p^{obs}$ is computed as the percentage of occurrences of each label. The
percentage in the $\ket{0}_n$ label is set to $0$, and the rest of $p^{obs}$
is rescaled. The activation ratio of a gene is the percentage of cells
expressing that gene on $\boldsymbol{X^b}$. (\textbf{B}) The parameters in
the $L_{enc}$ layer are initialized such that, independently, each qubit has
a probability of observing $1$ equal to the corresponding activation ratio,
and the parameters in $L_0$, $L_1$, $\cdots$, $L_n-1$ are initialized to zero.
In the matrix representation of $\boldsymbol{\theta}$, the diagonal values
belong to $L_{enc}$ layer, and the non-diagonal values to the regulation
layers. (\textbf{C}) The classical optimizer fits the smoothed distributions
$\hat{p}^{out}$ into $\hat{p}^{obs}$ using a gradient descent based algorithm
with the KL-divergence as the loss function and the mean square as the error
metric. (\textbf{D}) The matrix representation of the qscGRN model is used as
the adjacency matrix of the biological scGRN. In the adjacency matrix,
parameters smaller than $0.087$ are dropped (left side), assuming the
interaction is not strong enough to report. Thus, the network representation
of the biological scGRN (right side) is constructed using the remaining values
in the adjacency matrix.}
\label{figure3}
\end{figure}

\subsubsection{Recovery of Gene Regulatory Network
(FIGURE \ref{figure3}D)} We use the values of parameter
$\boldsymbol{\theta}$ of qscGRN model to construct the adjacency matrix of the
biological scGRN, as described in the matrix representation for qscGRN model
step. In the adjacency matrix, parameters with an absolute value less than
$0.087$ are set to $0$ because no significant rotation is performed by the
corresponding c-$R_y$ gate. The heatmap representation of the adjacency matrix
is shown on the left side of FIGURE \ref{figure3}D, where rows and
columns represent control and targetgenes, respectively, and strong regulation
relationships are highlighted. Finally, we construct the signed, directed,
weighted network representation (right side of FIGURE \ref{figure3}D)
of the biological scGRN using the adjacency matrix.

%% file: methods/data_sets.tex
\subsection{Data sets}
\textbf{LCL data set:} The scRNA-seq data was generated from lymphoblastoid
cell lines (LCLs), which are widely used cell model systems derived from human
primary B cells. Information about the experimental handling and acquisition of
data is provided in reference to our original study \cite{osorio2019single}.
The data set has been deposited in the Gene Expression Omnibus (GEO) database
with the accession number GSE126321. For this study, we merged this data set
with another LCL scRNA-seq data set \cite{sorelle2021single}, for which the
gene-barcode matrix files were downloaded from the GEO database using the
accession number GSE158275. The raw data matrix was pre-processed using
scGEAToolbox \cite{cai2020scgeatoolbox}. The processed matrix of $9,905$ genes
and $28,208$ lymphoblastoid cells was then transformed using the Pearson
residuals normalization \cite{lause2021analytic}. Then, expression values
of six genes (IRF4, REL, PAX5, RELA, PRDM1, AICDA) of the NF-$\kappa$B
signaling pathway were extracted. The $6$-gene expression matrix
($6$-by-$28,208$) was used as the input of the qscGRN analysis of this study.
The known regulatory relationships between genes were
obtained from the previously established B-cell differentiation circuit
model \cite{roy2019regulatory, sciammas2011incoherent}.

%% file: results/real_data.tex
\subsection{Real data LCL}
\begin{figure}[htbp]
\centerline{
    \includegraphics[width=0.6\textwidth,trim=0 10 0 3, clip]{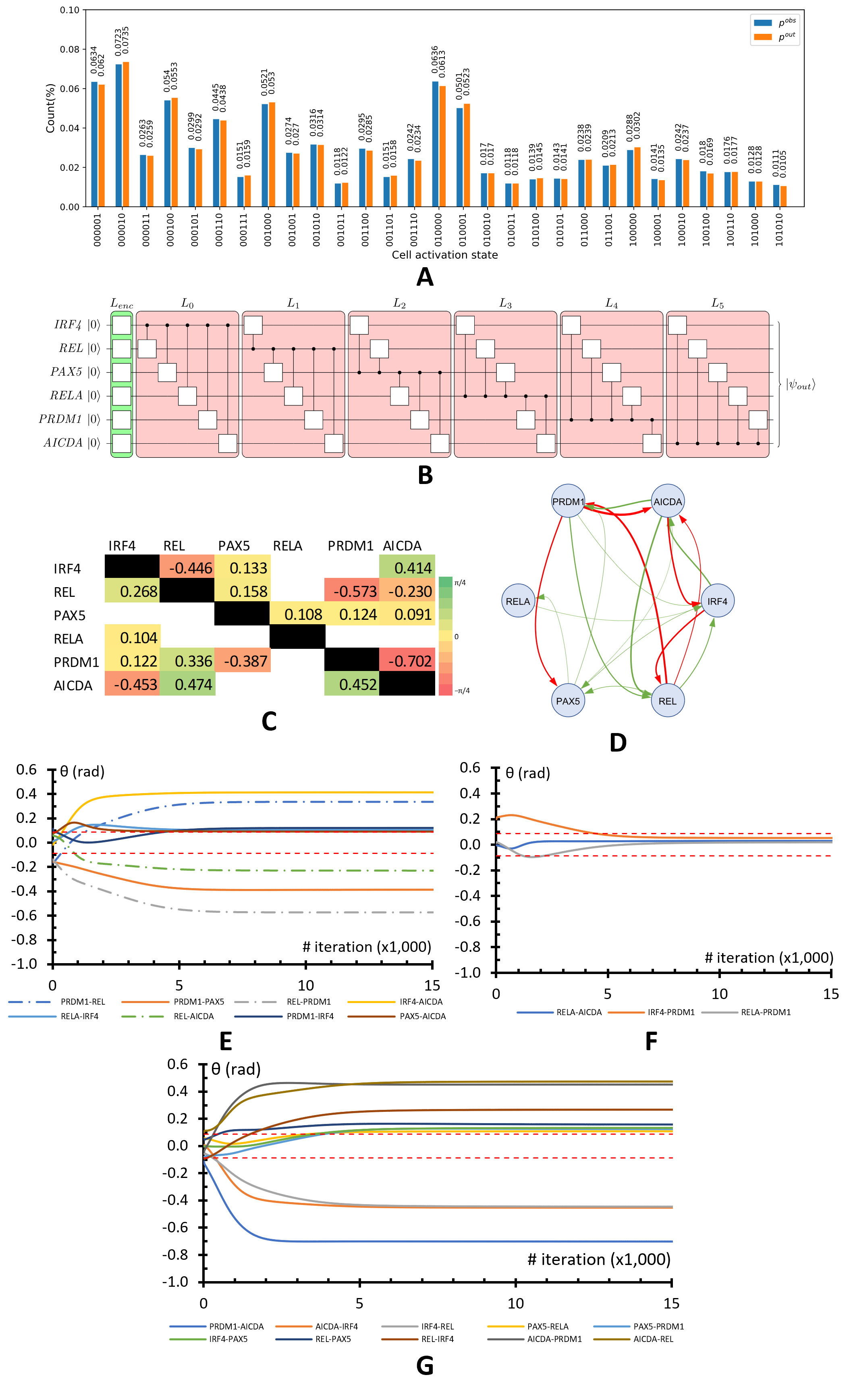}
}
\caption{Application of the qscGRN modeling with real scRNA-seq data.
(\textbf{A}) The observed and output distributions ($p^{obs}$ and $p^{out}$)
colored in blue and orange, respectively. $p^{obs}$ is computed using the
transformed scRNA-seq data. $p^{out}$ is obtained by measuring the output
register in the qscGRN model when the optimization is completed. (\textbf{B})
The schematic for a $6$-genes qscGRN model consists of a $L_{enc}$ layer and
six regulation layers that map relationships between genes in the quantum
framework. (\textbf{C}) The adjacency matrix of the biological scGRN, in which
parameters with an absolute value less than $0.087$ are removed. The heatmap
reveals the strength and direction of the interaction for a gene-gene
interaction. The diagonal elements are colored in black due to these
parameters are not trained during optimization. (\textbf{D}) A
directed-weighted representation of the biological scGRN recovered from the
quantum circuit. Up-regulation and down-regulation relationships are colored
in green and red, respectively. The thickness of each edge is proportional to
the absolute value of the corresponding parameter in the adjacency matrix.
(\textbf{E}) Evolution of parameters for gene-gene interaction that were
recovered from the quantum framework and are present in the NF-$\kappa$B
network during the first $15,000$ iterations. (\textbf{F}) Evolution of
parameters for gene-gene interactions that were not recovered but are present
in the NF-$\kappa$B network during the first $15,000$ iterations. (\textbf{G})
Evolution of parameters for gene-gene interaction that are predicted by the
quantum framework during the first $15,000$ iterations.}
\label{figure4}
\end{figure}

We used our quantum-classical framework to build the qscGRN model—a fully
connected quantum circuit—and compute the observed distribution $p^{obs}$
using the data set described in subsection \ref{pobs}. The input scRNA-seq expression matrix
contained more than $28,000$ cells belonging to the same cell type, the
lymphoblastoid cell. The $p^{obs}$ distribution maps the population of the
state in the scRNA-seq data into a vector space ($p^{obs}$ is represented in
blue—FIGURE \ref{figure4}A). The qscGRN model schema for the data set is a
$6$-qubit system and consists of an encoder layer and six regulation layers
(FIGURE \ref{figure4}B). We measured the output register of the qscGRN model to
recover the output distribution $p^{out}$ from the quantum framework. To
improve visualization, FIGURE \ref{figure4}A only shows the states with a
probability greater than $0.01$ due to the large size of the vector space.

Then, we optimized the parameter $\boldsymbol{\theta}$ in the qscGRN model
for $50,000$ iterations by minimizing the loss function $L(\boldsymbol{\theta})$
and using the smoothed distributions for $p^{out}$ and $p^{obs}$. Therefore,
the distribution $\hat{p}^{out}$ is fitted into $\hat{p}^{obs}$ during the
optimization, as shown in FIGURE \ref{figure4}A that visually proves the
similarity of both distributions after optimization. The $p^{out}$ after
optimization is represented in orange—FIGURE \ref{figure4}A. The similarity is
quantified using the loss function and error metrics that reached values of
$8.03 \times 10^{-4}$ and $3.04 \times 10{-5}$, respectively.

The value of the parameter $\boldsymbol{\theta}$ after optimization retrieves
an adjacency matrix (FIGURE \ref{figure4}C) that is used to construct the
biological scGRN. Then, we constructed a weighted, directed network from the
quantum framework using the non-diagonal elements of $\boldsymbol{\theta}$, as
shown in FIGURE \ref{figure4}D. We compared the sign of the element of each pair
of genes with the corresponding regulatory effect in the previously published
network, i.e., the baseline GRN \cite{roy2019regulatory, sciammas2011incoherent}
. The comparison results were measured using 3 metrics: accuracy, f1 score,
and precision, to quantify the prediction performance of the classical-quantum
framework. The qscGRN model recovers gene-gene relationships with an accuracy,
f1 score, and precision of $0.63$, $0.72$, and $0.78$, respectively.

FIGURE \ref{figure4}E shows the evolution of parameters for $8$ gene-gene pairs
(a control-target pair) in the qscGRN model during the optimization. These
pairs are relationships recovered from the quantum framework and are present
in the NF-$\kappa$B network. Gene pairs correctly recovered are represented
using a solid line, and pairs incorrectly recovered in long-dash-dot lines.
The $8$ gene-gene pairs almost reach their optimized value by $10,000$
iterations. Specifically, pairs such as IRF4-AICDA, PAX5-AICDA, and PRDM1-PAX5
that have reached their optimized value earlier than others are strongly
supported by the quantum framework. IRF4 is known to induce AICDA expression
through an indirect mechanism in the NF-$\kappa$B signaling cascade
\cite{sciammas2006graded}. PAX5 is suggested to be a player in the
B-lineage-specific control of AICDA transcription in a previous study
\cite{yadav2006identification}. PRDM1 is a master regulator that represses
PAX5 expression in B cells \cite{boi2015prdm1}.

FIGURE \ref{figure4}F shows $3$ gene-gene pairs with regulatory relationships are
present in the published baseline GRN but are removed by our qscGRN estimation.
The exclusion of these gene pairs suggests that the regulatory relationships
between them might be through indirect links. The baseline model failed to
distinguish between indirect and direct links. For example, one of the removed
gene pairs is IRF4-PRDM1 which has a parameter larger than $0.1$ during the
first $1,000$ iterations. The parameter value decreases gradually to close to
zero after $5,000$ iterations. The dropping suggests that IRF4-PRDM1’s
regulatory relationship might be through a third-party modulator. Indeed, IRF4
is known to inhibit BCL6 expression, and because BCL6 can repress PRDM1
\cite{tunyaplin2004direct, shapiro2005regulation}, it has been formally
speculated that the effects of IRF4 on PRDM1 expression might have been
mediated through inhibition of BCL6 expression \cite{teng2007irf4}.

Ten novel regulatory relationships between genes, which are not in the
published baseline GRN, were predicted by the qscGRN model when the computation
nearly reached their optimized value by $10,000$ iterations
(FIGURE \ref{figure4}G). We found that, for at least two of these newly discovered
regulatory relationships, there is experimental evidence supporting the
involved gene pairs. The first gene pair is IRF4-PAX5, for which our model
predicted that PAX5 is positively regulated by IRF4. Indeed, in a previous
study of the phosphoinositide-3 kinase signaling in B cells
\cite{abdelrasoul2018pi3k}, an experiment searching for transcription factors
that are activated by FOXO1 revealed that IRF4 is a potential candidate for
PAX5 regulation. The second gene pair is PRDM1-AICDA. PRDM1 has been found to
be able to silence the expression of AICDA, probably in a dose-dependent
manner \cite{nutt2011genetic}. These results suggest that our qscGRN method
recovered regulatory relationships that were missed in the published baseline
model.

%% file: discussion.tex
Finding ways to apply quantum computing in biological research is an active
research area \cite{prospect_qc_molecular, marx2021biology, emani2021quantum,
cheng2020application}. Many questions in biology can benefit from quantum
computing by exploring many possible parallel computational paths, but
identifying such questions remains challenging. Especially understanding how
to exploit quantum computers for progress in solving important biological
questions is crucial. The latest development of scRNA-seq has made it possible
to gather the transcriptome information from tens of thousands of individual
cells in a high-throughput manner. These complex data sets with unprecedented
detail are driving the development of new computational and statistical tools
that are revolutionizing our understanding of cellular processes. However,
quantum computation has not yet received enough attention in the face of this
single-cell big data revolution. As a consequence, we present our qscGRN method
for modeling interactions between genes to derive the quantum computing
framework for constructing GRNs. Below we discuss several aspects of
application issues.

\subsection{Mapping of scRNA-seq data in the quantum framework}
Typically, a correlation-based method obtains enough information from a
scRNA-seq data set to infer a GRN with a large number of genes. The gene-gene
interaction is calculated as a single value (summary statistic) using the
expression values of the available cells. On the other hand, our quantum
approach for GRN inference models a small number of genes due to the vector
space size, which is equal to the number of basis states, increases
exponentially with the number of genes. In other words, the number of cells
in binarized scRNA-seq data might be mapped to a moderate number of basis
states such that each basis state is mapped from the biological data. For
example, a $15$-qubit qscGRN model requires $2^{15}=32,768$ basis states;
however, a scRNA-seq data set with roughly $32,768$ cells would retrieve an
observed distribution without meaningful biological information mapped to
the basis states. The latest scRNA-seq technology has the capacity to allow
the transcriptome of millions of cells to be measured. To obtain enough cells,
we may also integrate multiple scRNA-seq data sets from the same cell types or
similar biological sources and perform statistical correction to remove the
batch effect. We can select informative genes such as highly variable genes
\cite{osorio2019singleexp} in advance, then proceed with our quantum-classical
pipeline. Thus, reducing the burden of a large number of genes in the model
while maintaining the biological relevance.

\subsection{Initialization values for the parameter
\texorpdfstring{$\boldsymbol{\theta}$}{Theta} in the qscGRN model}
The initialization of the parameter $\boldsymbol{\theta}$ determines the
starting point in the landscape of the loss function, thus indirectly setting
the difficulty for the optimization due to barren plateaus—which is an issue
when optimizing a parameterized quantum circuit. In subsection \ref{sectiontheta}
Initialization of the parameter $\boldsymbol{\theta}$ in the qscGRN, an
all-zeros approach is taken for the parameters in the c-$R_y$ gates.
Additionally, $2$ more initialization approaches for the c-$R_y$ gates were
tried using a random initialization with uniform and normal distributions.
The $3$ methods initialize the parameter $\boldsymbol{\theta}$ at $3$ different
positions in the landscape; however, only the all-zeros approach defines the
same point when running the workflow again. Thus, the all-zeros approach would
recover a biological scGRN consistently from the quantum framework since the
gradients are computed at the same starting point. Finally, the all-zeros
approach recovered $8$ gene-gene interactions with biological support, which
is larger than the other approaches.

\subsection{Advantages of the qscGRN over correlation- and
regression-based GRNs}
Correlation and regression-based methods are the most widely used methods for
GRN inference due to their computational efficiency. These methods typically
compute a correlation or regression coefficient for a gene pair using the total
number of cells in the data. The issue with correlation and regression methods
is that they rely on summary statistics. The relationship between the two genes
is measured using a single value of the summary statistics: correlation or
regression coefficient. Once computed, the coefficient becomes independent of
the total number of cells. Moreover, increasing the number of cells would not
substantially change the correlation and regression coefficients. Thus,
information in the scRNA-seq data is not fully used. The other issue is that
the coefficient is computed only between the two focal genes, regardless of
the expression values of other genes in the same biological system. The effect
of not considering other genes in the computation can result in a biased
coefficient, which does not represent the true behavior of the interaction.
There are methods such as partial correlation \cite{kim2015ppcor}, principal
component regression \cite{osorio2020sctenifoldnet}, and LASSO \cite{lassoref}
that may correct this. But the correcting effect is limited given the
all-to-all interactions cannot be modeled.

In contrast, our qscGRN method is based on the quantum framework, which uses
the Hilbert space and maps the binarized expression of each cell for $n$ genes
at once. The mapping of $n$ genes to the Hilbert space allows manipulating the
superposition of basis states in an $n$ qubit system and fitting the output
distribution to the observed distribution in the binarized scRNA-seq data.
Thus, biological information passes through the quantum circuit and is encoded
in a superposition state. In other words, a gene-gene relationship is computed
using information from an $n$ gene biological system all at once in the
quantum framework, which is an improvement compared to correlation-based
methods.